\documentclass[12pt]{iopart}   
\usepackage{graphicx}
\usepackage{amssymb}
\usepackage{citesort}
\begin{document}
\title[The Paired-Electron Crystal in the 2D Frustrated 
$\frac{1}{4}$-Filled Band]{The Paired-Electron Crystal in the 
Two-Dimensional Frustrated Quarter-Filled Band}
\author{H. Li$^1$, R. T. Clay$^2$, and S. Mazumdar$^1$}
\address{$^1$Department of Physics, University of Arizona, Tucson, AZ 85721}
\address{$^2$Department of Physics and Astronomy and HPC$^2$ Center for 
Computational Sciences, Mississippi State University, Mississippi State 
MS 39762}
\date{\today}
\begin{abstract}
The competition between antiferromagnetic and spin-singlet ground
states within quantum spin models and the $\frac{1}{2}$-filled band
Hubbard model has received intense scrutiny.  Here we demonstrate a
frustration-induced transition from N\'{e}el antiferromagnetism to
spin-singlet in the interacting $\frac{1}{4}$-filled band on an
anisotropic triangular lattice.  While the antiferromagnetic state has
equal charge densities 0.5 on all sites, the spin-singlet state is a
paired-electron crystal, with pairs of charge-rich sites separated by
pairs of charge-poor sites. The paired-electron crystal provides a
natural description of the spin-gapped state proximate to
superconductivity in many organic charge-transfer solids.
Pressure-induced superconductivity in these correlated-electron
systems is likely a transition from the $\frac{1}{4}$-filled band
valence bond solid to a valence bond liquid.
\end{abstract}

 \submitto{\JPCM}
\maketitle Quasi-two-dimensionality, strong electron-electron
repulsion, and proximity of the superconducting state to
semiconducting states with spatial broken symmetry are the common
features between superconducting cuprates and organic charge transfer
solids (CTS). Superconductivity in the CTS is reached from the
semiconducting state by the application of pressure at constant
carrier density. In the $\kappa$-(BEDT-TTF)$_2$X,
($\kappa$-(ET)$_2$X), the CTS family with the highest T$_c$, dimers of
BEDT-TTF (ET) molecules form an anisotropic triangular lattice.
Carrier concentration of one hole per dimer and commensurate
antiferromagnetism \cite{Kanoda06a} prompted theoretical description
of these CTS within the $\frac{1}{2}$-filled band Hubbard model
\cite{Kino95a,Kanoda06a} at ambient pressure. Within mean-field
theories, pressure enhances spin frustration, destroys
antiferromagnetism, and leads to superconductivity over a range of
anisotropy (see reference \cite{Kontani08a} for a review). This
viewpoint has been challenged by numerical calculations that
demonstrate the absence of superconductivity within the
triangular-lattice $\frac{1}{2}$-filled band Hubbard model for all $U$
and anisotropy \cite{Mizusaki06a,Clay08a,Tocchio09a}.  Recent
experiments also find a puzzling array of broken symmetries {\it
  different from the antiferromagnetic order} in the semiconducting
state proximate to superconductivity in the CTS, including spin-gap
\cite{Mori06a,Tajima06a,Shimizu07a,Yamashita09a}, charge-ordering
\cite{Tajima06a,Mori06a}, and coexisting charge-ordering and spin-gap
\cite{Tajima06a,Tamura06a}.  The close relationships between the
molecular and crystal structures of the two-dimensional (2D) CTS with
different semiconducting phases indicate the need for a common
theoretical description that can explain this panoply of competing and
coexisting orders.  We present here such a unified theoretical
description: we show that a single structural parameter, - {\it the
  extent of geometrical lattice frustration} - determines the nature
of the dominant broken symmetry.  We develop the concept of the
paired-electron crystal (PEC), which provides a new paradigm for
spin-singlet formation in dimensionality $> 1$, and perhaps also the
insight to understanding unconventional superconductivity in the CTS
and related $\frac{1}{4}$-filled band inorganic materials.  We point
out that concept of a PEC driven by intra-pair quantum exchange
occurring at intermediate densities in the electron gas has previously
been postulated \cite{Moulopoulos92a}. The mechanism of PEC formation
that we demonstrate here is related, with the difference that our
results apply to a lattice, with specific carrier concentration.

The idea that frustration-induced quantum effects can drive a
transition from antiferromagnetism to a resonating valence bond
spin-singlet state \cite{Anderson73a} has led to many exciting
developments in the theory of correlated-electron systems. With few
exceptions \cite{Poilblanc07a}, however, the consequences of lattice
frustration have been investigated mostly for the $\frac{1}{2}$-filled
band Hubbard model, which for large $U$ reduces to an
antiferromagnetic spin Hamiltonian. The effective $\frac{1}{2}$-filled
band model for the 2D CTS precludes charge order and clearly cannot be
the basis of the unifying theoretical description we seek. The number
of carriers $n$ per {\it molecule} of the CTS is $\frac{1}{2}$. Here
we focus on this specific $n$ and investigate the consequences of
lattice frustration within an electronic model with {\it simultaneous}
charge- and spin-frustration.

\begin{figure}
\centerline{\resizebox{2.5in}{!}{\includegraphics{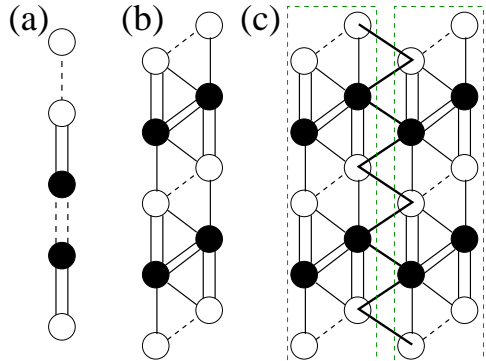}}}
\caption{(color online) (a) The PEC in the one dimensional
  $\frac{1}{4}$-filled band spin-Peierls state. Here, and in the
  following, filled (empty) circles correspond to sites with a charge
  density of 0.5+$\delta$ (0.5-$\delta$). Double lines are intra-dimer
  bonds; the inter-dimer bonds alternate between strong
  (double-dotted) and weak (dotted). The double-dotted bond is a
  localized singlet \cite{Clay03a}. (b) PEC state in the zigzag ladder
  lattice. Singlets are formed between the two chains \cite{Clay05a}.
  (c) Conceptualization of the 2D PEC emerging from the coupling of
  zigzag ladders.}
\label{cartoons}
\end{figure}

Before proceeding to 2D, it is instructive to summarize the behavior
of the $n=\frac{1}{2}$ strongly correlated band in one- and quasi
one-dimension, (1D and quasi-1D), where also quantum effects are
strong.  In 1D chains, a metal-insulator transition occurs at
intermediate temperatures to either a bond-dimerized state with equal
site charge densities, or a charge-ordered Wigner crystal with equal
intermolecular bond distances \cite{Clay03a}.  For strong enough
electron-phonon interactions, and not too strong nearest-neighbor
electron-electron repulsion, a spin-Peierls transition occurs within
both the insulating phases.  The spin-Peierls state can be thought of
as further bond-dimerization of the bond-dimerized state with
schematic charge occupancies $\cdots 1100 \cdots$ \cite{Clay03a},
where `1' and `0' denote charge-rich and charge-poor sites.  The
interdimer 1-1 bond, stronger than the 0-0 bond, constitutes a
localized singlet bond (see Fig.~\ref{cartoons}(a)).  This structure
is the simplest example of the PEC and is observed in many 1D CTS
\cite{Clay03a}.  Coexistence of charge-order and spin-gap is a
consequence of co-operation between electron-phonon and
nearest-neighbor antiferromagnetic spin interactions.  This quantum
effect dominates for a wide range of parameters over the classical
effect due to the nearest-neighbor Coulomb repulsion, which prefers
the Wigner crystal configuration $\cdots1010\cdots$ \cite{Clay03a}.

There also exist ``ladder'' CTS with isolated pairs of $n=\frac{1}{2}$
chains \cite{Rovira00a}.  The low temperature spin-gap state found in
these quasi-1D materials can be explained as a PEC in a two-leg zigzag
ladder lattice, with {\it interchain} localized singlet bonds (see
Fig.~\ref{cartoons}(b)) The PEC occurs when the frustrating interchain
hopping exceeds a critical value \cite{Clay05a}.  Interestingly, in
both the $n=\frac{1}{2}$ 1D chain and the zigzag ladder, the PECs can
be thought to be derived from the corresponding $n=1$ valence bond
solids by simply replacing alternate spin-singlet bonds with pairs of
vacancies.  Coupling $n=1$ zigzag ladders to generate the triangular
lattice necessarily leads to the loss of the valence bond solid
structure \cite{Anderson73a}.  As shown in Fig.~\ref{cartoons}(c),
however, similar coupling of the $n=\frac{1}{2}$ zigzag ladders can in
principle lead to a 2D PEC with localized spin-singlet bonds. We
demonstrate below that this is exactly what happens in 2D for large
enough lattice frustration. Furthermore, the PEC with localized
singlet valence bonds between the charge-rich sites necessarily has a
spin-gap, and thus a transition occurs from the antiferromagnetic to
the spin-gapped state in $\frac{1}{4}$-filled 2D systems with
increasing lattice frustration.
\begin{figure}
\centerline{\resizebox{4.0in}{!}{\includegraphics{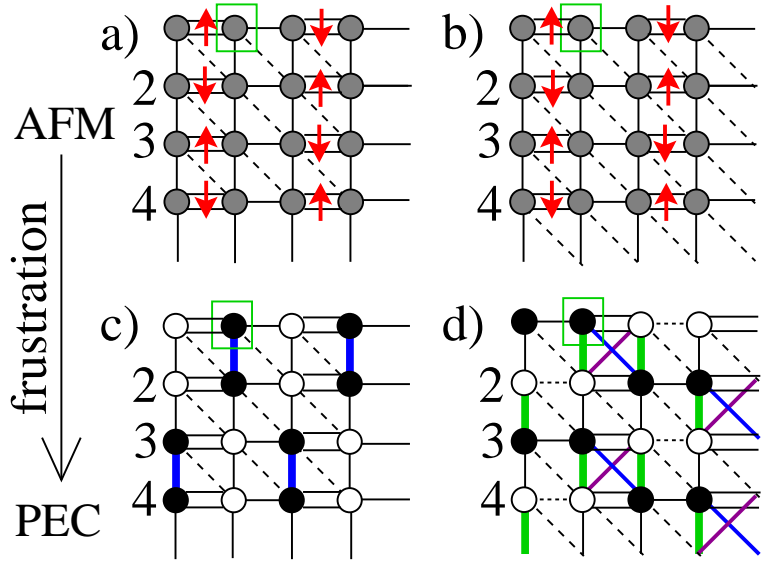}}}
\caption{(color online) (a) OBC and (b) PBC 4$\times$4 lattices for
  $t^{\prime}<t^{\prime}_c$. Double bonds and thick lines imply strong
  bonds; thin lines imply weak bonds. The dashed lines are the
  diagonal bonds whose strength is varied.  Charge densities are
  uniform as indicated by grey circles and spin ordering corresponds
  to antiferromagnetism (AFM).  (c) and (d) show the PEC state
  occuring for $t^{\prime}>t^{\prime}_c$.  Black and white circles
  represent charge-rich and charge-poor sites.  Singlet bonds form
  between the charge-rich sites in the PEC.  Numbers correspond to the
  chain indices in Fig.~\ref{data}(a)-(b).  Boxes mark site 2 of chain
  1. Spin-spin correlations between this site and sites on other
  chains are shown in Fig.~\ref{data}.}
\label{lattices}
\end{figure}

We consider the $\frac{1}{4}$-filled extended Hubbard model on an
anisotropic triangular lattice,
\begin{eqnarray}
H=-\sum_{\nu,\langle ij\rangle_\nu}t_\nu(1+\alpha_\nu\Delta_{ij})B_{ij} 
+\frac{1}{2}\sum_{\nu,\langle ij\rangle_\nu} K^\nu_\alpha \Delta_{ij}^2 \label{ham} \\
+\beta \sum_i v_i n_i + \frac{K_\beta}{2} \sum_i v_i^2  
+ U\sum_i n_{i\uparrow}n_{i\downarrow} + 
\frac{1}{2}\sum_{\langle ij\rangle}V_{ij} n_i n_j \nonumber.
\end{eqnarray}

In Eq.~\ref{ham}, $\nu$ runs over three lattice directions
($\nu=x,y,x-y$, see Fig.~\ref{lattices}). We have considered $0.5 \leq
t_y \leq t_x$, but report results here for $t_x=t_y=t$ only.  In what
follows, all energies are in units of $t$. Our calculations are for
frustrated lattices $0 \le t_{x-y}\equiv t^\prime \le 1$. Our results
for negative $t^\prime$ are qualitatively similar and we do not report
these separately.  $B_{ij}=\sum_\sigma(c^\dagger_{i\sigma}c_{j\sigma}+
H.c.)$ is the electron hopping between sites $i$ and $j$ with electron
creation (annihilation) operators $c^\dagger_{i\sigma}$
($c_{i\sigma}$).  $\alpha_\nu$ is the inter-site electron-phonon
coupling, $K^\nu_\alpha$ is the corresponding spring constant, and
$\Delta_{ij}$ is the distortion of the $i$--$j$ bond, determined
self-consistently \cite{Clay03a}.  $v_i$ is the intra-site phonon
coordinate and $\beta$ is the intra-site electron-phonon coupling with
corresponding spring constant $K_\beta$. $U$ and $V_{ij}$ are on-site
and nearest-neighbor Coulomb interactions, respectively.

Given the complexity and detailed nature of our numerical results, we
present the outcome of our calculations at the outset.  We start with
the square lattice ($t^\prime=0$, $V_{x-y}=0$) limit of Eq.~\ref{ham},
where two different semiconducting states are possible: (i) an
in-phase dimerized state with all site charge densities equal, - such
a state with one electron per dimer unit cell is a 2D antiferromagnet
(see Fig.~\ref{lattices}(a),(b)); (ii) and for sufficiently large
$V_x$ and $V_y$, the Wigner crystal with checkerboard site
occupancies.  Note that the dimerized antiferromagnetic phase and the
Wigner crystal are mutually exclusive.  Based on the experimentally
observed antiferromagnetism in weakly frustrated CTS and the
observation that in all such cases the lattice is dimerized, we choose
parameters that give antiferromagnetism in the square lattice
limit. In
\begin{figure}
\centerline{\resizebox{4.0in}{!}{\includegraphics{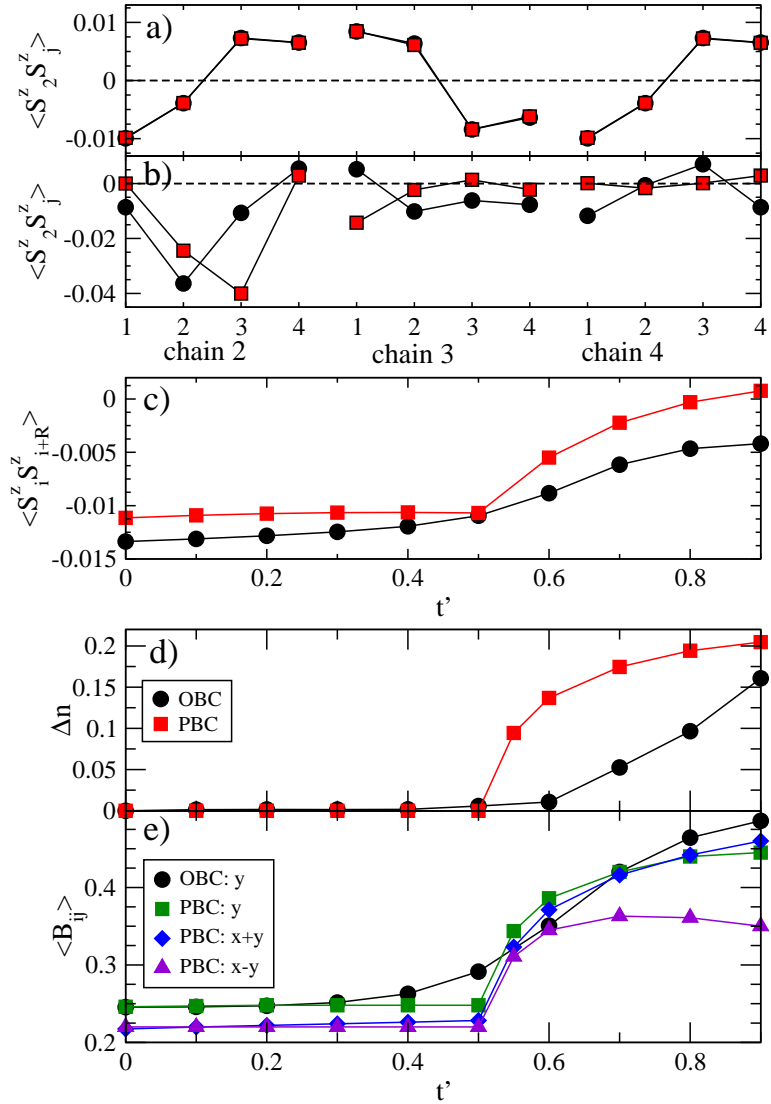}}}
\caption{(color online) (a) Z-Z spin-spin correlations between site 2
  in chain 1 (marked with a box in Fig.~\ref{lattices}), and sites 1 -
  4 in chains 2, 3 and 4 for $t^{\prime}=0$.  The chain indexing is
  shown in Fig.~\ref{lattices}.  In all panels, (a)-(e), circles and
  squares correspond to OBC and PBC calculations, respectively.  (b)
  Spin-spin correlations as in (a), but with $t^\prime=0.7$.  In
  panels (c)-(e), we plot several order parameters as a function of
  $t^\prime$.  (c) Spin-spin correlations between sites that are
  members of the most distant dimers.  (d) Difference in
  charge-densities between charge-rich and charge-poor sites. (e)
  Bond-orders between pairs of nearest-neighbor sites forming
  localized spin-singlets.  For the OBC lattice, these bonds are along
  the $y$ direction (Fig.~\ref{lattices}(c)).  For the PBC lattice,
  bonds along $y$, $x+y$, and $x-y$ all change at the PEC transition
  (see Fig.~\ref{lattices}(d)); we plot these using squares, diamonds,
  and triangles, respectively.}
\label{data}
\end{figure}
Fig.~\ref{lattices} we present results of exact ground state
calculations on 4$\times$4 lattices with two different boundary
conditions, open (OBC) and periodic (PBC).  Both are periodic along
$x$ and $y$ directions.  The OBC is however open along the $x-y$
direction with 12 $t^\prime$ bonds. We did not explicitly include the
electron-phonon interactions in this case. Rather, we work in the
limit $\alpha_\nu=\beta=0$; in order to have antiferromagnetic order
at $t^\prime=0$ here we incorporate intrinsic dimerization $t_x=t\pm
\delta_t$ as indicated in Fig.~\ref{lattices}(a). We keep $t_y=t$
fixed and gradually increase $t^\prime$ from zero and monitor
different order parameters that indicate the antiferromagnetism-to-PEC
transition (see below).  In contrast to the OBC, the PBC is periodic
along $x-y$ and has 16 $t^\prime$ bonds.  We now work with nonzero
$\alpha_\nu$ and $\beta$, and all bond distortions $\Delta_{ij}$ and
intra-site electron-phonon coordinates $v_i$ are obtained
self-consistently \cite{Clay03a}. Once again we keep $t_y$ fixed and
increase $t^\prime$ from zero, but now calculate all bond distortions,
charge densities and spin-spin correlations
self-consistently. Importantly, the bond alternation along the
$x$-axis, as seen in Fig.~\ref{lattices}(b), and the resultant
antiferromagnetism, appear now as a consequence of the
self-consistent calculation.

In Fig.~\ref{data} we give numerical data for OBC and PBC lattices
with Eq.~\ref{ham} parameters $U=4$, $V_x=V_y=1, V_{x-y}=0$. For the
OBC lattice $\delta_t=0.2$ and $\alpha_\nu=\beta=0$, and for the PBC
$\alpha_x=1.3$, $\alpha_y=1.0$, $K^x_\alpha=K^y_\alpha=2$,
$\beta=0.1$, and $K_\beta=2$.  Our calculations indicate that the
charge-densities are uniform for both the OBC and PBC lattices for
$t^\prime < t^\prime_c$, where the lattices are antiferromagnetic.
For $t^\prime > t^\prime_c$, the charge densities are nonuniform and
the antiferromagnetism has disappeared. Spin-singlet formation is
indicated by the large increases in the strengths of the vertical bond
indicated by the thick lines in Fig.~\ref{lattices}(c), and of the
diagonal and vertical bonds indicated by the thick lines in
Fig.~\ref{lattices}(d), respectively.  Below we give the details of
our calculations.

In Fig.~\ref{data}(a) we plot the z-z spin-spin correlation functions
$\langle S_2^zS_j^z \rangle$ for $t^\prime=0$ between fixed site 2 of
chain 1 (marked with box on first row of each lattice in
Fig.~\ref{lattices}) and sites $j$, labeled sequentially 1, 2, 3, 4
from the left, on neighboring chains labeled 2, 3, 4 in
Fig.~\ref{lattices}.  In Fig.~\ref{data}(a) only, the average
spin-spin correlation with each chain has been shifted to zero in
order to clearly show the antiferromagnetic pattern, which is $\cdots-
- + + \cdots$ and $\cdots+ + - - \cdots$, indicating N\'eel ordering
of the dimer spin moments in both lattices.  The loss of this pattern
in Fig.~\ref{data}(b) for large $t^\prime=0.7$ indicates loss of
antiferromagnetic order.  In Fig.~ \ref{data}(c) we plot the z-z
spin-spin correlation $\langle S_i^zS_{i+R}^z\rangle$ between
maximally separated dimers at location $i$ and $i+R$, which measures
the strength of the antiferromagnetic moment.  This correlation is
nearly constant and negative (as expected for the antiferromagnetic
ground state, see Fig.~\ref{lattices}(a) and (b)) until
$t^\prime_c\sim 0.5$, beyond which the antiferromagnetic order is
destroyed.

We define $\Delta n$ as the charge-density difference between
charge-rich and charge-poor sites. Fig.~\ref{data}(d) shows the rapid
increase in $\Delta n$, starting from zero, for $t^\prime>t^\prime_c$
with both lattices.  Simultaneously with charge order, there occurs a
jump in the bond orders $\langle B_{ij} \rangle$ between the sites
that form the localized spin-singlets. This is shown in
Fig.~\ref{data}(e).  These bond orders are by far the strongest in
both lattices for $t^\prime > t^\prime_c$.  The spin-spin correlation
between the same pairs of sites becomes strongly negative at the same
$t^{\prime}$, even as all other spin-spin correlations approach zero
(Fig.~\ref{data}(b)), indicating spin-singlet character of the
strongest bonds.  Taken together, the results of Figs.~\ref{data} give
conclusive evidence for the antiferromagnetism-to-PEC transition shown
in Fig.~\ref{lattices}.

We have obtained the same antiferromagnetism-to-PEC transition for
$0<U \leq 8$, $0 \leq V_x,V_y,V_{x-y} \leq 2$ with both $V_x=V_y$ and
$V_x \neq V_y$, $0.5 \leq t_y \leq t_x$.  The Coulomb interaction
parameters thus cover a broad range appropriate for 2D CTS
\cite{Kino95a,Clay02a,Clay03a}.  As we discuss below, the hopping
parameters are also relevant for 2D CTS with strong frustration.  Here
we emphasize that although our Hamiltonian has many different
parameters, in each case we fix all parameters corresponding to an
antiferromagnetic state, and vary a single parameter $t^\prime$ to get
the transition to the PEC. {\it Thus in all cases the transition is
  driven by the frustration alone.}

We performed several checks on our calculations to verify that the
antiferromagnetism-to-PEC transition we find is not the result of
finite-size effects. Possible finite size effects are usually related
to (a) large discrete energy gaps in the excitation spectrum, or (b)
degeneracies in the electron occupancy schemes in the $U=V_{ij}=0$
limit. Had our results been a consequence of (a), the energy gaps
would have shown a sudden change at $t^\prime_c$, and this should have
been visible also in the $U=V_{ij}=0$ limit. The energy gaps between
the one-electron levels of our lattices change monotonically as a
function of $t^\prime$, and there is no sudden change at $t^\prime_c$;
we therefore discount possibility (a).  The one-electron occupancy
scheme is degenerate for all $t^\prime$ for the PBC, and nondegenerate
for all $0<t^\prime<1$ for the OBC. The {\it identical} results
obtained with the PBC and OBC, in spite of this difference, as well as
the {\it same} one-electron degeneracies for both
$t^\prime<t^\prime_c$ and $t^\prime>t^\prime_c$, indicate that the
transition is not associated with lattice distortions peculiar to
finite systems (such as Jahn-Teller distortions). While we have shown
here the results for two different boundary conditions only, we have
found the same antiferromagnetism-to-PEC transition in yet another
lattice, with open boundaries in both $y$ and $x-y$ directions. The
complete absence of band structure-related effects, identical results
obtained with three different boundary conditions, and the wide range
of parameters over which we found the antiferromagnetism-to-PEC
transition, taken together, give us confidence that the transition we
observe is real.

We believe that there is a relatively simple physical reason for the
antiferromagnet-to-PEC transition in the $\frac{1}{4}$-filled band in
2D.  As already suggested in \cite{Anderson73a}, and discussed
extensively also in the recent literature \cite{Balents10a}, geometric
frustration strongly enhances the tendency to form spin-singlet
valence bonds. In the dimerized $\frac{1}{4}$-filled band
antiferromagnet such singlet bonds will occur between dimer unit cells
(see Fig.~\ref{lattices}). Focusing on individual sites
within the dimer units, strong singlet bonds must occur between
nearest neighbors while bonds between distant sites must be weaker,
which implies a wavefunction that is a superposition of valence bond
diagrams with strong bonds between nearest neighbor sites and diagrams
with weaker bonds between distant sites.  Such a structure in the
$\frac{1}{4}$-filled band will necessarily have charge-ordering
accompanied by bond distortion (see Fig.~1(c)).  Yet another way to
understand the exceptional stability of the PEC at $n=\frac{1}{2}$ is
to recognize this as a commensurability effect.  Recall that
transitions driven by electron-electron interactions, from the
metallic to the Mott-Hubbard semiconductor at $n=1$, or to the Wigner
crystal in bipartite lattices at $n=\frac{1}{2}$, are essentially
consequences of high order commensurability.  The PEC with the numbers
of bonded and vacant pairs of sites exactly equal at $n=\frac{1}{2}$
is also commensurate, and there is gain in exchange energy from the
strong bonding between the charge-rich sites.

The PEC paradigm enables us to explain the seemingly widely different
behavior in the semiconducting states of CTS that undergo transition
to superconductivity under pressure.  Although 2D CTS occur in
different crystalline forms, it has been pointed out that they can
generally be described as anisotropic triangular lattices with varying
degrees of frustration
\cite{Kino95a,Mori99a,Clay02a,Hotta03a,Fukuyama06a,Mori06a}.  The PEC
structures in all cases are arrived at simply by noting that the
charge order pattern is (i) $\cdots1100\cdots$ along two of three
directions
\begin{figure}
\centerline{\resizebox{5.0in}{!}{\includegraphics{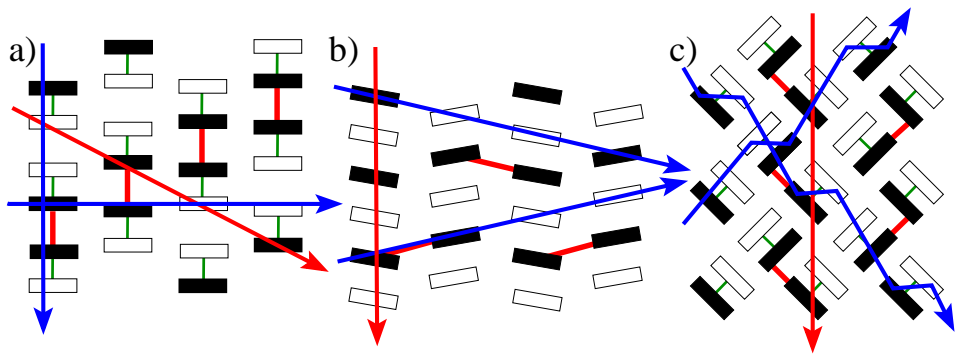}}}
\caption{(color online) PEC structures for (a)
  EtMe$_3$P[Pd(dmit)$_2$]$_2$ \cite{Tamura06a}, and (b)
  $\theta$-(ET)$_2$RbZn(SCN)$_4$ \cite{Watanabe07a}.  Black and white
  boxes are charge-rich and charge-poor molecules.  Thin bonds are
  strong intradimer hopping integrals and thick bonds are spin-singlet
  bonds.  (c) Charge occupancies and singlet bonds for
  $\kappa$-(ET)$_2$Cu$_2$(CN)$_3$ in the $T=0$ limit, as suggested
  from extension of the PEC concept to this system.  It is conceivable
  that the PEC here is short-range at finite temperatures.  The arrows
  in all cases indicate charge carrier paths $\cdots1100\cdots$ and
  $\cdots1010\cdots$.}
\label{cts}
\end{figure}
of the triangular lattice, and (ii) $\cdots1010\cdots$ in the
direction of weakest hopping (see Fig.~\ref{cartoons}(c),
Fig.~\ref{lattices}(c) and (d).) In Fig.~\ref{cts}, we give the PEC
patterns for $\beta^{\prime}$-X[Pd(dmit)$_2$]$_2$, $\theta$-(ET)$_2$X
and $\kappa$-(ET)$_2$X.  PECs for other crystal structures can be
obtained by simple extrapolations. Thus, for example, the crystal
structures of $\beta$, $\beta^{\prime}$ and $\beta^{\prime\prime}$
salts are related \cite{Mori98c,Hotta03a,Beta}; similarly, the
$\alpha$-phase is obtained by slight modifications of the basic
$\theta$ structure motif \cite{Mori99a,Hotta03a,Fukuyama06a,Mori06a}.
We discuss below the individual PEC structures of Fig.~\ref{cts} and
specific systems that exhibit transitions to the superconducting state
from a charge-ordered or a spin-gapped state (we also discuss the
antiferromagnetism-to-superconductivity case below).  We have focused
only on ground states here. The temperatures T$_{SG}$ at which the
spin gap opens will depend on detailed lattice structures, and could
be vanishingly small in some materials.

(i) The family of X[Pd(dmit)$_2$]$_2$, where X is a monovalent cation,
have the $\beta^{\prime}$ crystal structure.  Systems with relatively
weak frustration exhibit antiferromagnetism \cite{Tamura09a}, but
those with lattice structures closest to isotropic triangular,
X=Et$_2$Me$_2$Sb, and X=EtMe$_3$P exhibit spin gaps \cite{Itou08a}.
Thus $t^{\prime}$ in these systems lie in the region where
theoretically we find the PEC to dominate over the antiferromagnetism.
In Fig.~\ref{cts}(a) we have shown our proposed PEC pattern for the
semiconducting phase of
$\beta^{\prime}$-EtMe$_3$P[Pd(dmit)$_2$]$_2$. Note that this pattern
is obtained by simple rotations of the PECs of Fig.~\ref{lattices},
with the y-axis (x-axis) of Fig.~\ref{lattices}(c)
(Fig.~\ref{lattices}(d)) corresponding to the intrastack axis of
Fig.~\ref{cts}(a).  The period 4 intrastack site charge densities {\it
  and} intermolecular distances (strong intradimer 1-0 bond, and
alternating weak interdimer 1-1 bond and weaker 0-0 bond) of
Fig.~\ref{cts}(a) have all been observed in X=EtMe$_3$P
\cite{Tamura05a,Tamura06a}. In our model calculation, we treat
molecules as point objects, but the orientation of molecules (both
intra- and inter-layer) strongly affects the magnitude of the spin
gap. Clearly, ``parallel'' orientation of molecules, as occurs in
X=EtMe$_3$P (see Fig.~3b in Reference \cite{Tamura06a}) makes the
formation of a strong spin-singlet bond easier, and enhances the
tendency to transition to the PEC self-consistently. The favorable
orientations are responsible for the high T$_{SG} \sim 25$ K here.

Like $\beta^{\prime}$-EtMe$_3$P[Pd(dmit)$_2$]$_2$, the salt
$\beta$-($meso$-DMET)$_2$PF$_6$ also exhibits a pressure-induced
transition from charge order to superconductivity. The charge order
here also has patterns $\cdots1100\cdots$ in two directions and
$\cdots1010\cdots$ along the third direction, with almost the same PEC
structure (see Fig.~2 in \cite{Kimura06a}.)

(ii) The PEC of Fig.~\ref{cts}(b) explains the spin-gap phase in the
$\theta$-(ET)$_2$MM$^{\prime}$(SCN)$_4$ family \cite{Mori06a}.  The
charge order and bond patterns here are again obtained from rotations
of the PECs in Fig.~\ref{lattices}, with the y-axis (x-axis) and the
diagonal direction of Fig.~\ref{lattices}c (d) corresponding to the
strong diagonal hoppings in the $\theta$-lattice of Fig.~\ref{cts}(b).
Although theoretical \cite{Clay02a} and experimental
\cite{Watanabe04a} studies showed that the charge order corresponds to
the ``horizontal stripe'' of charge-rich sites in Fig.~\ref{cts}(b),
the mechanism of the spin-gap transition was not understood until now.
The ``zigzag'' horizontal stripe here (see Fig.~\ref{cts}(b)) is part
of a 2D lattice and a simple one dimensional spin-Peierls mechanism
for the spin-gap is not satisfactory.  In the $\theta$-lattice the
weakest hopping is along the stack, and has the charge pattern
$\cdots1010\cdots$. The weak intrastack hopping is about half the
strong interstack hoppings \cite{Mori99a}, making the $\theta$-lattice
strongly frustrated already at high temperatures. With decreasing
temperature, the intrastack lattice parameter decreases sharply
relative to the interstack lattice parameters \cite{Watanabe07a}.
Within our theory, the lattice contraction leads to increased
intrastack hopping, and therefore increased frustration and transition
to the PEC with spin-gap within our theory.

As has been noted above, the lattice structures of $\theta$- and
$\alpha$-salts are related \cite{Mori99a,Hotta03a,Fukuyama06a}. The
small differences between the hopping integrals in the two classes
have significant effect on band structures calculated for uncorrelated
charge carriers, but are of less significance in the presence of
strong electron-electron interactions. Frustration is therefore
expected to play a strong role also in the $\alpha$-(ET)$_2$X.  The
observation that the state obtained below the metal-insulator
transition temperature in $\alpha$-(ET)$_2$I$_3$ exhibits coexisting
charge-order and spin-gap \cite{Rothaemel86a}, as opposed to
antiferromagnetism, confirms this conjecture.  The recent observation
of phason-like modes in the charge-ordered spin-gapped state in this
material confirms that the PEC state involves coupled charge, bond,
and spin degrees of freedom \cite{Ivek10a}.

(iii) To date, the $\kappa$-(ET)$_2$X have been modeled as a
triangular lattice of dimers of ET molecules. However, it is possible
to consider the lattice also as a triangular lattice of {\it monomers}
with hoppings between each molecule and six neighbors (see e.g. Fig.~9
in \cite{Mori99a}).  There is a fundamental similarity between the
$\kappa$- and $\theta$-structures, in that the two directions defined
by the interdimer hopping integrals $p$ in \cite{Mori99a} once again
correspond to the $x$ and $y$ directions in Fig~\ref{lattices}, with
the hopping integral labeled $b2$ in \cite{Mori99a} taking the role of
$t^{\prime}$ in our calculations.  The main difference from the
$\theta$-structure is the very strong dimerization which enhances the
tendency to antiferromagnetism. Indeed, simultaneous strong
dimerization and frustration makes the $\kappa$-(ET)$_2$X closer to
$\beta^{\prime}$-EtMe$_3$P[Pd(dmit)$_2$]$_2$ than to the other
ET-salts \cite{KappaNote,Kandpal09a,Nakamura09a}.

In $\kappa$-(ET)$_2$X relative orientations of neighboring dimers are
nearly perpendicular. Assuming comparable electron-phonon couplings in
the $\kappa$-(ET)$_2$X and other CTS, much stronger lattice distortion
would be needed in the $\kappa$-(ET)$_2$X to form static singlet
bonds, and any spin-gap would be very small. This is probably the
reason why a direct antiferromagnetism-to-superconductivity transition
occurs under pressure in most $\kappa$-(ET)$_2$X.  The nearly
isotropic $\kappa$-(ET)$_2$Cu$_2$(CN)$_3$ does not exhibit
antiferromagnetism and until recently was considered to be a spin
liquid \cite{Shimizu03a}. Specific heat \cite{Yamashita08a} and
thermal conductivity \cite{Yamashita09a} measurements have claimed
contradictory evidences for gapless versus gapped energy spectrum here
\cite{Yamashita08a,Yamashita09a}.  Both experiments also find evidence
for a possible phase transition at 6 K.  More recently, a second-order
thermodynamic phase transition near 6 K that is accompanied by strong
anisotropic lattice effects has been demonstrated.
\cite{Manna10a}. These investigators have determined that the
resultant entropy change cannot be accounted for in terms of spin
degrees of freedom alone, and have suggested that {\it charge degrees
  of freedom play a strong in the transition \cite{Manna10a}}.
Similarly, measurements of the dielectric response have shown
increasing and frequency-dependent dielectric constant below 60 K, and
possible antiferroelectric ordering of the dipoles at a transition
temperature of $\sim$ 9 K, which requires {\it unequal site charges on
  the dimer unit cell \cite{Abdel-Jawad10a}}.  Both of these latter
experiments \cite{Manna10a,Abdel-Jawad10a} are in agreement with the
PEC picture, although because of the relative orientations of the ET
molecules the PEC could be short-ranged.  In Fig.\ref{cts}(c) we have
shown the proposed PEC here: the pattern of the charge-ordering is
once again $\cdots1100\cdots$ along two paths and $\cdots1100\cdots$
along the third.  One possible pattern for forming spin-singlet bonds
is shown in Fig.\ref{cts}(c).  With short-ranged PEC order at finite
temperatures there would be tendency to form domain-walls, as
suggested from the dielectric constant measurements
\cite{Abdel-Jawad10a}. Importantly, (i) the charge order pattern
obtained using the PEC concept outlined above is {\it identical to
  those proposed in \cite{Abdel-Jawad10a,KappaCONote}}, and (ii) the
absence of similar charge-disproportionation in other
$\kappa$-(ET)$_2$X with smaller frustration is explained naturally
within our theory.

We conclude that the tendency to form the PEC is ubiquitous to the
strongly frustrated $\frac{1}{4}$-filled band.  Experimentally,
increased frustration induced by pressure or chemical substitution
causes antiferromagnetism-to-superconductivity \cite{Kanoda06a},
antiferromagnetism-to-PEC \cite{Tamura09a} or PEC-to-superconductivity
\cite{Kimura06a,Shimizu07a} transitions.  A unified theoretical
approach for strongly-correlated superconductivity in
$\frac{1}{4}$-filled materials follows if it is assumed that the
superconductivity is a consequence of the transition from the valence
bond solids of Fig.~\ref{cts} to valence bond liquids with mobile
spin-singlet bonds. The singlet bonds of the $\frac{1}{4}$-filled band
PEC can be visualized as effective negative-$U$ centers in a
$\frac{1}{2}$-filled band \cite{Mazumdar08a}.  We have recently shown
that within such an effective $\frac{1}{2}$-filled band negative-$U$
model with repulsive interaction between the on-site pairs there
occurs a first-order transition from a charge-ordered state to
superconductivity with increased frustration \cite{Mazumdar08a}.  The
PEC concept lies at the interface of theories emphasizing
electron-electron and electron-phonon interactions: the bipolarons in
the PEC are bound not by extraordinarily strong electron-phonon
interactions \cite{Micnas90a,Alexandrov}, but by antiferromagnetic
correlations.  Direct transition to the superconducting state from
antiferromagnetism in most $\kappa$-systems is conceivably
energetically more favorable since this does not require static
lattice distortion, which we have pointed out can be energetically
expensive because of their crystal structures.

 Interestingly, there exist other unconventional superconductors in
 which electron-electron interactions, frustration and
 $\frac{1}{4}$-filling appear all appear to play significant roles,
 giving support to the mechanism of superconductivity proposed by us
 \cite{Mazumdar08a}.  The cobalt ions in the family of materials
 Na$_x$CoO$_2$, for example, form an isotropic triangular lattice, and
 electron-electron interactions in these are strong
 \cite{Limelette06a}.  Superconductivity occurs only in the hydrated
 material Na$_x$CoO$_2$ $\cdot$ $y$H$_2$O, with the H$_2$O layers
 occurring between the Na and Co layers. The Co valency in the
 hydrated superconductor has been determined to be very close to +3.5,
 indicating a $\frac{1}{4}$-filled Co hole-band ($\frac{3}{4}$-filled
 electron-band) \cite{Sakurai06a,Banobre-Lopez09a}.  Similarly,
 LiTi$_2$O$_4$ and CuRh$_2$S$_4$ are superconductors with highly
 frustrated 3D spinel structure and effective $\frac{1}{4}$-filled
 $d$-bands. As with the CTS, superconductivity in these spinels is
 proximate to a semiconductor, as seen from the unconventional
 metal-insulator transition in CuIr$_2$S$_4$ \cite{Radaelli02a}, which
 is isoelectronic with CuRh$_2$S$_4$.  The effective
 $\frac{1}{4}$-filled band of Ir$^{3.5+}$ ions undergoes an
 orbitally-driven Peierls instability at the metal-insulator
 transition \cite{Khomskii05a}, with charge-ordering
 Ir$^{4+}$-Ir$^{4+}$-Ir$^{3+}$-Ir$^{3+}$ (equivalent to the
 $\cdots1100\cdots$ pattern occuring in certain directions of the
 PEC), and {\it spin-singlet} Ir$^{4+}$-Ir$^{4+}$ bonds, in spite of
 the 3D nature of the crystal.  We believe that the charge-ordering of
 the 3D lattice here reflects the same tendency of the frustrating
 interacting $\frac{1}{4}$-filled band to form the PEC that we have
 demonstrated in 2D. Work is currently in progress to demonstrate the
 transition to superconductivity from the $\frac{1}{4}$-filled band
 PEC.  This work was supported by the US Department of Energy grant
 DE-FG02-06ER46315.


\begin{thebibliography}{10}
\bibitem{Beta}
The $\beta^\prime$ and related structures in the X[Pd(dmit)$_2$]$_2$ CTS are
  similar to but not identical to the $\beta^\prime$ structure in ET-based CTS.
\bibitem{KappaNote}
The frustration parameter $t^\prime/t$ commonly quoted for $\kappa$-(ET)$_2$X
  \cite{Kandpal09a,Nakamura09a} refers to the effective $t^\prime$ and $t$
  between dimers, and is not a useful measure of the degree of frustration in
  the lattice of monomers.
\bibitem{KappaCONote}
Fig.~4(c) is oriented as in \cite{Mori99a} and is rotated by 90$^o$ relative to
  the lattice shown in \cite{Abdel-Jawad10a}.
\end{thebibliography}
\end{document}